\documentclass[graybox]{svmult}

\usepackage{mathptmx}       
\usepackage{helvet}         
\usepackage{courier}        
\usepackage{type1cm}        
\usepackage{makeidx}         
\usepackage{graphicx}        
\usepackage{multicol}        
\usepackage[bottom]{footmisc}

\makeindex             

\begin{document}

\title{Chimera states in pulse coupled neural networks: \\
the influence of dilution and noise}
\author{Simona Olmi and Alessandro Torcini}
\institute{Simona Olmi \at
Aix Marseille Univ, Inserm, INS, Institut de Neurosciences des Syst{\' e}mes, Marseille, France;\\
CNR - Consiglio Nazionale delle Ricerche - Istituto dei Sistemi Complessi, 50019 Sesto Fiorentino, Italy;\\
INFN - Istituto Nazionale di Fisica Nucleare - Sezione di Firenze, 50019 Sesto Fiorentino, Italy.\\ 
\email{simona.olmi@univ-amu.fr}
\and Alessandro Torcini \at 
Laboratoire de Physique Th\'eorique et Mod\'elisation
Universit\'e de Cergy-Pontoise - CNRS, UMR 8089,
95302 Cergy-Pontoise cedex, France;
Aix Marseille Univ, Inserm, INMED, Institute de Neurobiologie
de la M\'editerran\'ee and INS, Institut de Neurosciences des Syst{\' e}mes, Marseille, France;\\
Aix-Marseille Universit\'{e}, Universit\'{e} de Toulon, CNRS, CPT, UMR 7332, 13288 Marseille, France;\\
CNR - Consiglio Nazionale delle Ricerche - Istituto dei Sistemi Complessi, 50019 Sesto Fiorentino, Italy.\\
\email{alessandro.torcini@u-cergy.fr}}
%
%
\maketitle

\abstract*{Each chapter should be preceded by an abstract (10--15 lines long) that summarizes the content. The abstract will appear \textit{online} at 
\url{www.SpringerLink.com} and be available with unrestricted access. This allows unregistered users to read the abstract as a teaser for the complete chapter. 
As a general rule the abstracts will not appear in the printed version of your book unless it is the style of your particular book or that of the series to which your book belongs.
Please use the 'starred' version of the new Springer \texttt{abstract} command for typesetting the text of the online abstracts (cf. source file of this chapter 
template \texttt{abstract}) and include them with the source files of your manuscript. Use the plain \texttt{abstract} command if the abstract is also 
to appear in the printed version of the book.}

\texttt{We analyse the possible dynamical states emerging for two symmetrically pulse coupled populations of leaky integrate-and-fire neurons. In particular, we observe broken symmetry states in this set-up: namely, breathing chimeras, where one population is fully synchronized and the other is 
in a state of partial synchronization (PS) as well as generalized chimera states,
where both populations are in PS, but with different levels of synchronization. 
Symmetric macroscopic states are also present, ranging from quasi-periodic motions, 
to collective chaos, from splay states to population anti-phase partial synchronization.
We then investigate the influence disorder, random link removal or noise, 
on the dynamics of collective solutions in this model. 
As a result, we observe that broken symmetry chimera-like states, with 
both populations partially synchronized, persist up to $80 \%$ of broken links
and up to noise amplitudes $\simeq 8 \%$ of threshold-reset distance.
Furthermore, the introduction of disorder on symmetric chaotic state has a constructive effect,
namely to induce the emergence of chimera-like states at intermediate dilution or noise level.}

\section{Introduction}
\label{sec:1}
The emergence of broken symmetry states ({\it Chimera states}) in population of oscillators or rotators is an extremely
popular and active research field nowdays, in particular after that experimental evidences for the existence
of these states have been reported in several contexts ranging from ensembles of mechanical oscillators, to laser dynamics, to populations of chemical oscillators 
(for a recent review on the subject see \cite{pan2015}).
Chimera states in neural systems have been firstly reported by Sakaguchi for a chain of nonlocally coupled Hodgkin-Huxley 
models with excitatory and inhibitory synaptic coupling~\cite{sakaguchi2006}, while the first evidence of chimeras
in models of globally coupled populations has been reported in~\cite{olmi2010} for leaky integrate-and-fire (LIF) excitatory neurons. More recently, 
chimeras in LIF networks have analysed in different contexts, ranging from small-network topology~\cite{lehnertz2014}, to chains of non-locally
coupled LIFs with refractoriness~\cite{hoevel2015}, to the emergence of chimeras in a single 
fully coupled neural population~\cite{bolotov2015}.
Globally pulse coupled Winfree models, reproducing $\theta$-neuron dynamics, also support
chimera states, ranging from breathing (periodic and quasi-periodic) to chaotic ones~\cite{pazo2014}.

Since the connectivity in the brain is definitely sparse and noise sources cannot
be avoided, it is fundamental in order to understand the possible relevance of
chimera-like states in neural dynamics to test for the robustness of these solutions to
dilution and to the presence of noise. Studies in this direction have been
performed mainly for oscillator models~\cite{laing2012,laing2012b,anna2016} or excitable systems \cite{semenova}.
In particular, chimera states in random diluted Erd\"os-Renyi networks have been observed
up to a dilution of $8 \%$ of the links~\cite{laing2012}, furthermore it has been
shown that noise has not only a washing out effects on chimera solutions, but 
it can also have a constructive role promoting new dynamical phenomena~\cite{laing2012b,semenova}.
 
In this paper, we focus on the dynamics of two fully pulse coupled populations of 
excitatory LIF neurons with stronger synaptic coupling among the neurons of the same
population and a weaker coupling with those of the other population, similarly to the 
simplest set-up showing the emergence of chimera states in phase oscillator networks~\cite{abrams}.
Furthermore, the neurons are synaptically connected via the transmission of pulses of finite
duration. This model for globally coupled systems reveal the emergence of broken
symmetry population states, chimera-like,~\cite{olmi2010}, as well as
of chimera states even within a single population~\cite{bolotov2015}. 

Our main aim is to study how the macroscopic solutions, found in the deterministic
fully coupled networks, will be modified by considering randomly connected networks 
of increasing dilution and by adding noise of increasing amplitude to the system.
In particular, after having introduced the considered models in Sect. 1.2,
we will report the complete phase diagram for the macroscopic solutions of the fully 
coupled case in Sect. 1.3.
These solutions vary from chimera-like, to symmetric solutions with complex
dynamic ranging from collective quasi-periodic dynamics to macroscopic chaos.
Furthermore, we will concentrate on the effect of random dilution and noise
on the dynamics of broken symmetry and chaotic states in Sect. 1.4 and 1.5.
Finally, we will devote Sect. 1.6 to a brief discussion of the reported results.
The algorithm employed to exactly integrate the fully coupled populations
is explained in the Appendix.

\section{The model}
\label{sec:2}

Firstly we consider two fully coupled networks, each made of $N$ LIF oscillators. Following Refs.~\cite{zillmer2}, the membrane potential $x_j^{(k)}(t)$
of the $j-th$ oscillator ($j=1,\ldots,N$) of the $k$th population ($k=0,1$) evolves according to the differential equation,
\begin{equation}
\label{eq:single}
\dot{x}_{j}^{(k)}(t)= a-x_{j}^{(k)}(t)+g_s E^{(k)}(t) +g_c E^{(1-k)}(t)
\end{equation}
where $a >1$ is the suprathreshold input current, while $g_s > 0$  and $g_c > 0$ gauge the self- and, resp., cross-coupling strength of the excitatory
interaction. 
The discharge mechanism
operating in real neurons is modeled by assuming that when the membrane potential reaches the threshold value $x_j^{(k)} = x_{th} = 1$, it
is reset to the value $x_j^{(k)} = x_R =0$, while a $\alpha$-pulse $p(t) = \alpha^2 t \exp{-\alpha t}$͒ is transmitted to and instantaneously received by 
the connected neurons. For this kind of pulses the field $E^{(k)}(t)$ generated by the neurons of the population $k$, satisfies the differential equation
\begin{equation}
\label{eq:E}
  \ddot E^{(k)}(t) +2\alpha\dot E^{(k)}(t)+\alpha^2 E^{(k)}(t)=\frac{\alpha^2}{N}\sum_{j,n} \delta(t-t_{j,n}^{(k)}) \ ,
\end{equation}
where $t_{j,n}^{(k)}$ is the $n$th spiking time of the $j$th neuron within the population $k$, and the sum is restricted to times smaller than $t$.
In the limit case $g_s=g_c=g$, the two populations can be seen as a single one made of $2N$ neurons with an effective coupling constant $G= 2g$.

Secondly we consider two random undirected Erd\"os-Renyi networks, each made of $N$ LIF oscillators and with
an average in-degree K, therefore the probability to have a link between two neuron is simply $K/N$.
We assume that the membrane potential $x_j^{(k)}(t)$ of the $j-th$ oscillator of the $k$th population 
($k=0,1$) evolves according to the differential equation
\begin{equation}
\label{eq:single:diluted}
\dot{x}_{j}^{(k)}(t)= a-x_{j}^{(k)}(t)+g_s E_j^{(k)}(t) +g_c {\overline E}^{(1-k)}(t) \quad;
\end{equation}
where the field $E_j^{(k)}(t)$ takes in account of the pulses received by neuron
$j$ from neurons of its own population, while the field ${\overline E}^{(1-k)}(t)$
represents the effect of the neuron beloging to the other population. 

In particular, $E_j^{(k)}(t)$ is the linear superposition of the pulses $p(t)$ received by neuron $i$ of the 
$k$th-population at all times $t_n<t$ (the integer index $n$ orders the sequence of the pulses emitted in the network), 
namely :
\begin{equation}
\label{eq:E:dil1}
 E_j^{(k)}(t)=\frac{1}{K} \sum_i \sum_{n| t_n<t} C_{i(n),j}^{(k)} \theta(t-t_n)p(t-t_n) \quad ,
\end{equation}
where $\theta(x)$ is the Heavyside function and the connectivity matrix $C_{i,j}^{(k)}$ has entries 1 (0) depending if the neuron $j$ 
presents a post-synaptic neuron connection with neuron $i$ or not. For each neuron we should
introduce a different field, since each neuron has a different connectivity in the network.
It is more convenient to turn also this time, as previously done for the globally coupled case, the explicit 
Eq.~(\ref{eq:E:dil1}) into the following differential equation
\begin{equation}
\label{eq:E:dil2}
  \ddot E_j^{(k)}(t) +2\alpha\dot E_j^{(k)}(t)+\alpha^2 E_j^{(k)}(t)= \frac{\alpha^2}{K}\sum_{i,n} C_{i(n),j}^{(k)}\delta(t-t_{i,n}^{(k)}) \ .
\end{equation}
Furthermore, ${\overline E}^{(1-k)}(t)=\frac{1}{N}\sum_{i=1}^N E_i^{(1-k)}(t)$ represents a ``mean field'' effect of the 
second population on the neuron of the first population, since it is the average of all the fields $E_i^{(1-k)}$ of the second population.
As a result, the dynamics of the neural network model takes the more ``canonical'' form of a set of coupled ordinary
differential Eqs. (\ref{eq:single:diluted}) and (\ref{eq:E:dil2}), which can be analyzed with the standard methods of dynamical systems. 
The setup we have employed, diluted random connectivity within each population, but
mean-field like cross coupling, will favour the stabilization of the broken symmetry
state as suggested in~\cite{laing2012}.
The have studied the diluted networks in so-called massively connected case, namely where  the average in-degree is proportional to the system size $K = (1-d) \times N$.

Finally we consider two diluted networks with noise. The noise is introduced in the system every time the membrane potential has reached the threshold value and it is
reset to the reset value. In particular, instead of using a reset value $x_R = 0$, the neuron is reset to a random value chosen in the interval $x_R \in [-\Delta, \Delta]$,
where $\Delta$ takes into account the level of the noise. In this case the percentage of dilution is kept fixed ($d=0.2$). 

The integration of the above models is performed exactly in terms of so-called event driven maps
analogously to what previously done in~\cite{zillmer2,olmi2010}, for the two fully coupled cases, where non trivial
round-off problems can occur a more refined event driven map has been developed and it is explained in the Appendix.

The degree of synchronization within each population of neurons can be quantified by introducing the typical order parameter used for phase oscillators
$r^{(k)}(t) =|\langle {\rm e}^{i \theta_j^{(k)}(t)}\rangle|$, where  $\theta_j^{(k)}$ is the phase of the $j$th oscillator, that can be properly
defined as a (suitably scaled) time variable~\cite{winfree}, $\theta_j^{(k)}(t) = 2\pi (t-t_{j,n}^{(k)})/(t-t_{m,n-1}^{(k)})$, where $n$
identifies the time of the last spike emitted by the $j$th neuron, while $m$ identifies the neuron that has emitted the last spike at time $t$. One can
verify that this phase is bounded between 0 and $2\pi$, as it should. Furthermore, the fully synchronized regime
corresponds to $r^{(k)} \equiv 1$, and in the asynchronous regime one expects $r^{(k)} \simeq 1/\sqrt{K}$, where $K$ is the average in-degree
of the network.


\section{Fully Coupled Network: Phase Diagram}
\label{subsec:2}

\begin{figure}
\begin{center}
\includegraphics*[angle=0,width=0.8\textwidth]{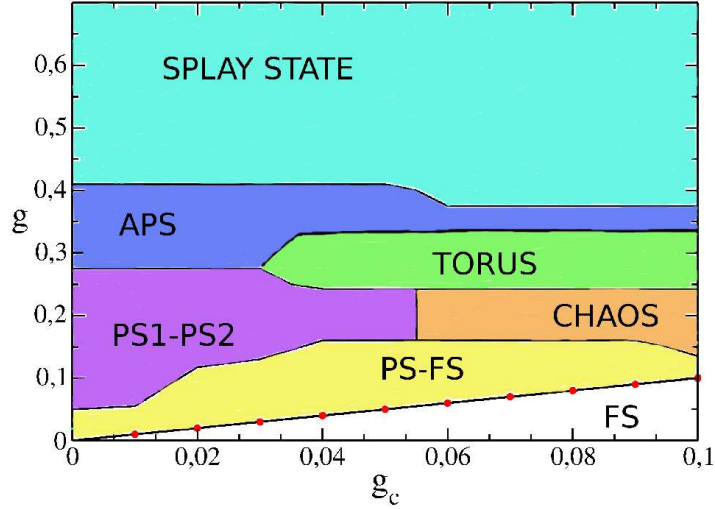}
\end{center}
\caption{(Color Online) Phase diagram in the $(g_c,g_s)$-plane reporting the 
stability region of the observed various collective solutions. For the
definitions of the different phases see the text.}
\label{fig.1}
\end{figure}

\begin{figure}
\begin{center}
\includegraphics*[angle=0,width=8.1cm]{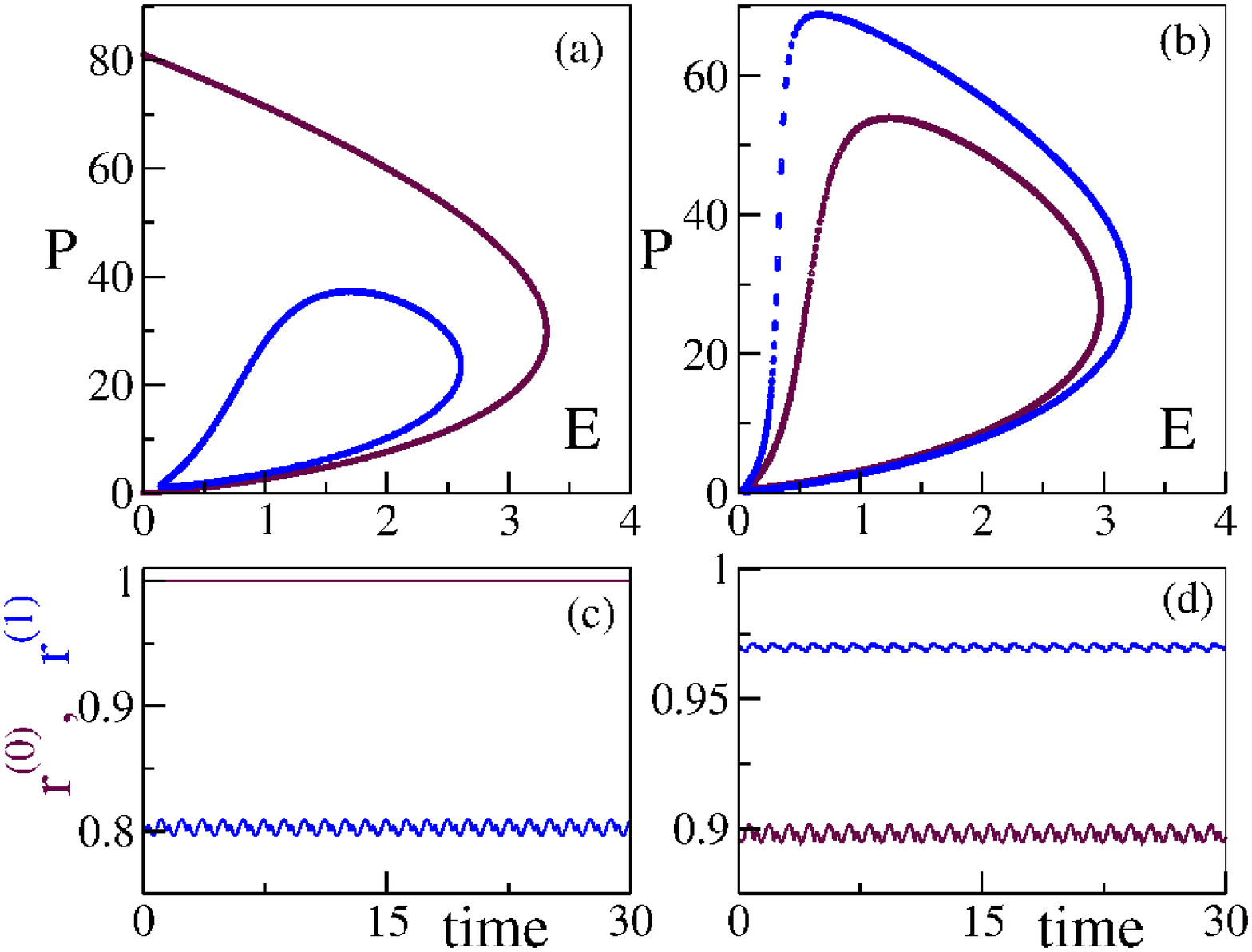}\\
\includegraphics*[angle=0,width=8cm]{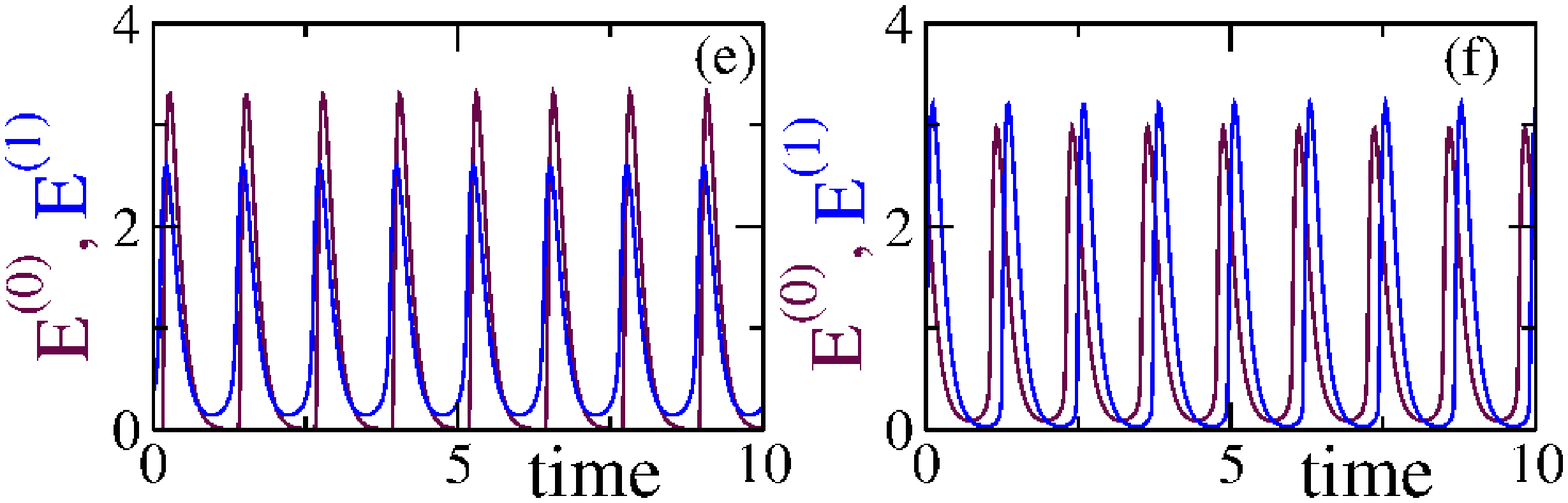}
\end{center}
\caption{(Color Online)  Macroscopic attractors displayed by reporting
$P \equiv E+\alpha \dot E$ vs $E$ for a PS-FS state (a) and 
a PS1-PS2 (b), the time evolution of the corresponding order 
parameters $r^{(0)}$ and $r^{(1)}$ is also reported in (c) and (d).
In panels (e), (f) are reported the time behaviors of the macroscopic fields
$E^{(0)}$ and $E^{(1)}$.
The variables corresponding to population 0 (resp. 1) are shown in blue (resp. maroon).
As regards the parameter values, $(g_c=0.07,g_s=0.1)$
in (a),(c) and $(g_c=0.02,g_s=0.17)$ in (b),(d).
}
\label{fig.1bis}
\end{figure}

The phase plane $(g_c,g_s)$ shown in Fig. ~\ref{fig.1} has been obtained by studying the model (\ref{eq:single},\ref{eq:E}) for $a=1.3$ and $\alpha=9$.
As already mentioned, along the diagonal ($g=g_s=g_c$) the two population model (\ref{eq:single}) reduces to a single population
with coupling strength $G=2g$. For our choice of $a$ and $\alpha$ values, the system exhibits 
{\it Partial Sybnchronization} PS, where the macroscopic field displays collective periodic oscillations
and the microscopic dynamics is quasiperiodic~\cite{vvres,zillmer2}.
Below the diagonal, the evolution is still symmetric but the neurons are now {\it Fully Synchronized} (FS); the neurons of both populations fire
at the unison. More interesting phenomena can be observed above the diagonal. In this situation a solution with broken
symmetry emerges naturally, where one population is FS and the other is PS, this represents a generalized form of chimera state.
In particular, one observes that while the order parameter of one population is exactly one, the other oscillates periodically,
as shown in Fig.~\ref{fig.1bis} (c). Therefore this chimera state can be classified as a {\it periodically breathing chimera}, which has been previously reported for 
the Kuramoto model~\cite{abrams,pikov_chimera} as well as for a two population network of rotators in \cite{olmi2015}.
Despite the macroscopic filds $E^{(0)}$ and $E^{1}$ are both oscillating periodically and locked, as evident from Figs.~\ref{fig.1bis} (a),(e),
the two populations are characterized by different behaviour at a microscopic level, where the neurons are periodic in the FS population and 
quasi-periodic in the PS population. This means that the neurons subject to two different
linear combinations of $E^{(0)}$ and $E^{(1)}$ behave differently: a population locks with the forcing field, while the other one behaves quasi-periodically.

Another even more interesting symmetry broken state (termed PS1-PS2) can be observed for 
larger $g_s$-values and $ g_c < 0.055$; in this case both populations exhibit PS, 
but their dynamics take place over two different attractors with two different degrees of synchronization, as shown in Fig.~\ref{fig.1bis} (b),(d). 
Analogously to the PS-FS state, the two fields are periodic and phase locked, as it can be seen by looking at the time behavior of $E^{(0)}$ and
$E^{(1)}$ in Fig. \ref{fig.1bis} (f). However, at variance with PS-FS, here both populations exhibit quasi-periodic motions.
This simmetry broken state can be also considered a Chimera state and it has been reported only for LIF populations so-far~\cite{olmi2010}.

\begin{figure}
\includegraphics*[angle=0,width=0.66\textwidth]{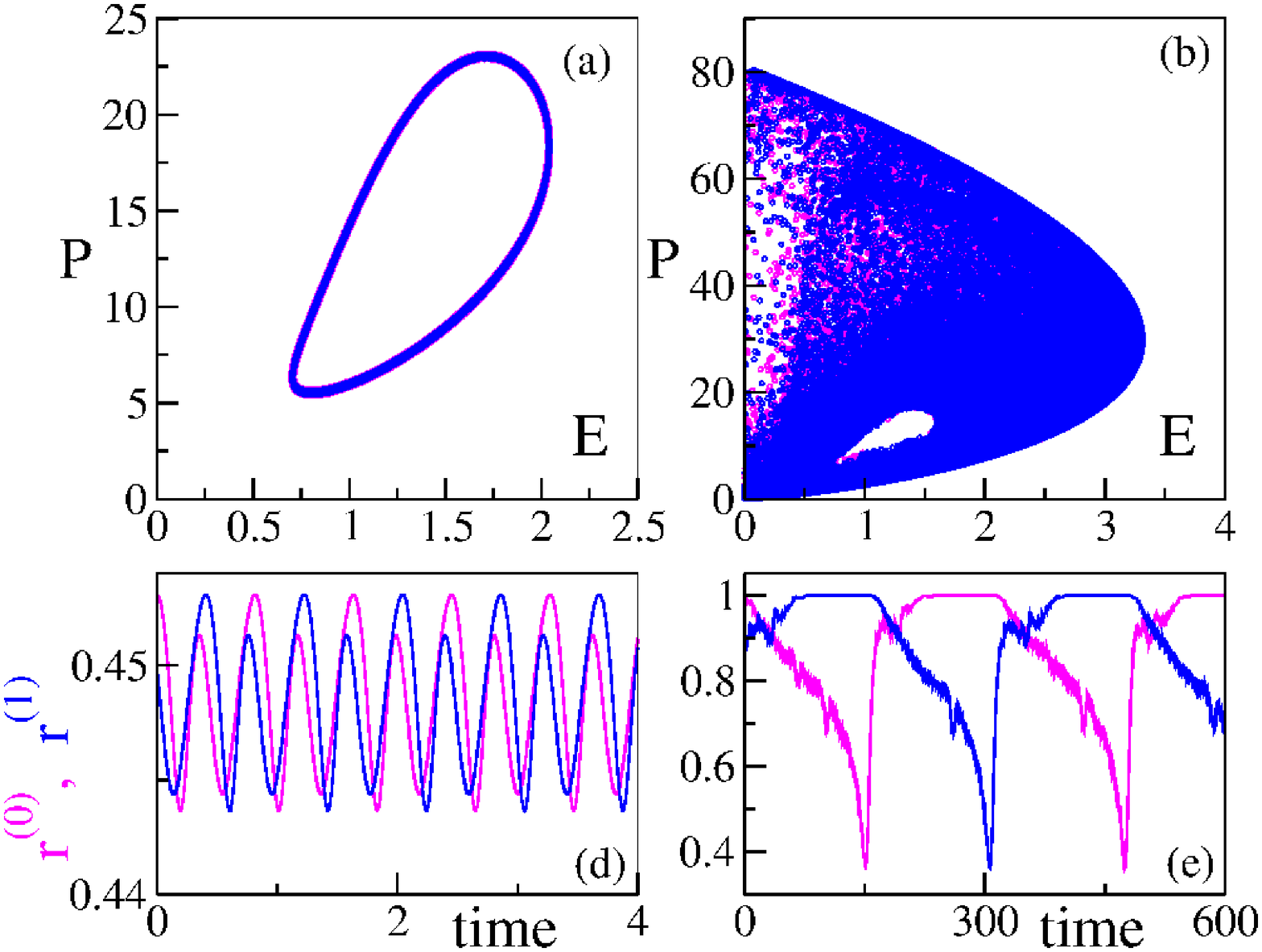}
\includegraphics*[angle=0,width=0.33\textwidth]{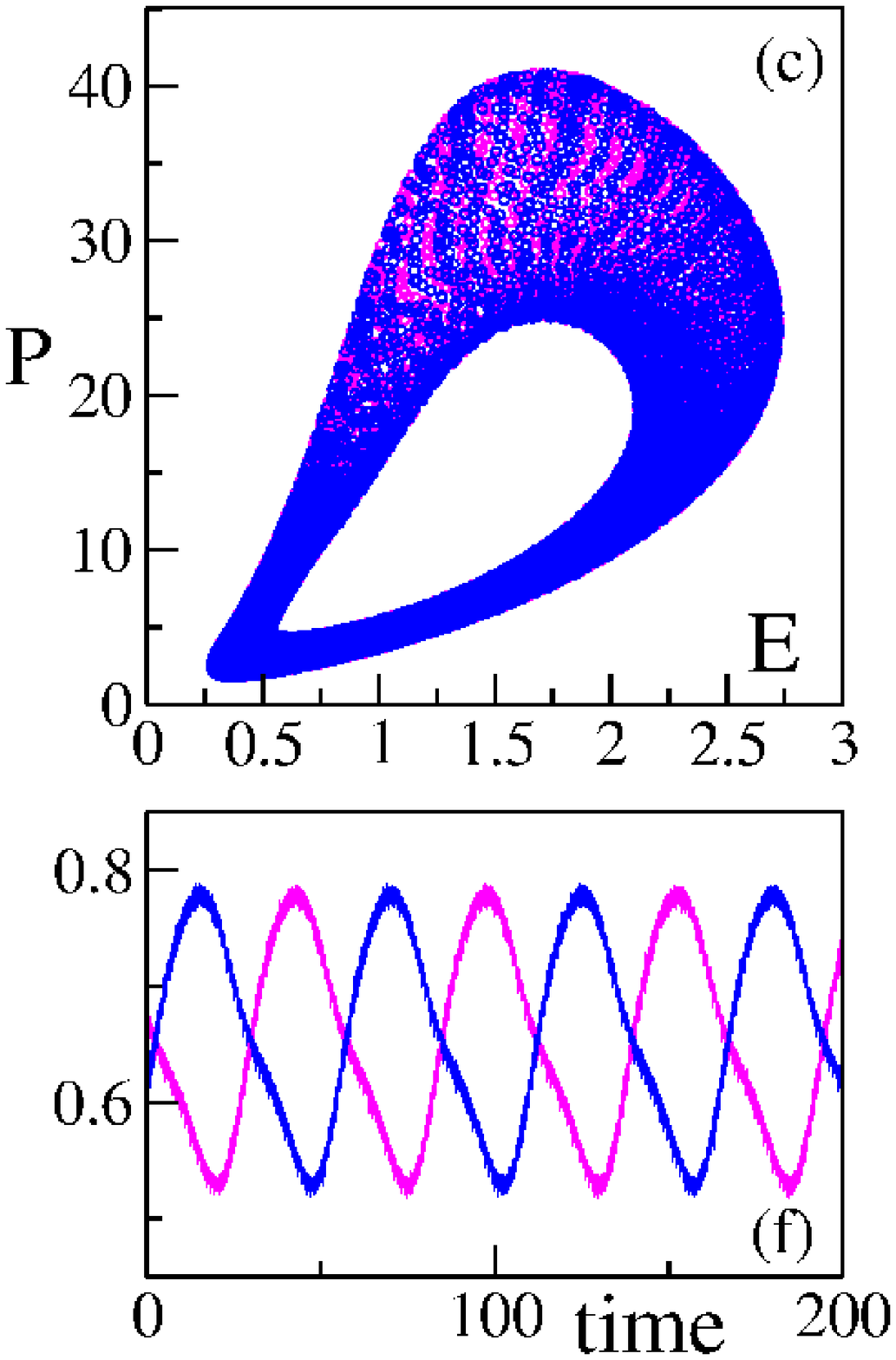}
\includegraphics*[angle=0,width=1\textwidth]{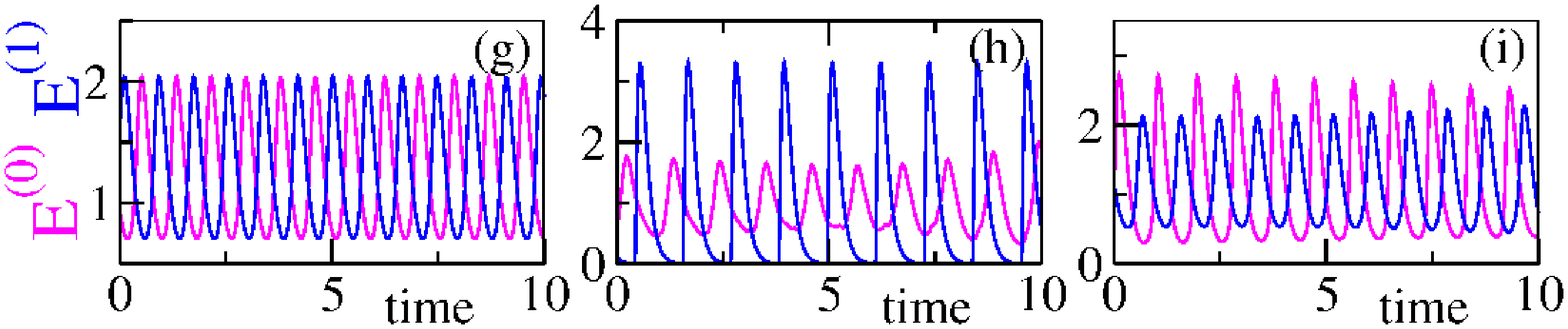}
\caption{(Color Online) Macroscopic attractors displayed by reporting
$P \equiv E+\alpha \dot E$ vs $E$ for an APS (a), a chaotic state (b), and a TORUS state (c),
the time evolution of the corresponding order parameters $r^{(0)}$ and $r^{(1)}$ is also reported in (d), (e) and (f). 
In panels (g), (h), (i) are reported the time behaviors of the macroscopic fields $E^{(0)}$ and $E^{(1)}$.
The variables corresponding to population 0 (resp. 1) are shown in magenta (resp. blue). As regards the parameter values, $(g_c=0.07,g_s=0.35)$
in (a),(d),(g), $(g_c=0.08,g_s=0.16)$  in (b),(e),(h) and  $(g_c=0.07,g_s=0.3)$ in (c),(f),(i).
}
\label{fig.2}
\end{figure}

For larger $g_s$ values the symmetry between the two collective fields is recovered with the only difference of phase shift between
the two fields which oscillate in antiphase and this is why we term this regime {\it Antiphase Partial Synchronization} (APS) (see Fig. \ref{fig.2} (a), (g)). 
In this regime, at finite $N$ the istantaneous maximum Lyapunov exponent strongly fluctuates and we cannot exclude that this regime is
{\it weakly chaotic}. Analogously to the chaotic behaviour found in single population of massively coupled LIFs~\cite{simo},
we expect that the chaoticity disappears in the thermodynamic limit.
However, it is peculiar the behaviour of the order parameters in this case, as shown in Fig. \ref{fig.2} (d): the two populations are not
equally synchronized and the two order parameters $r^{(0)}$ and $r^{(1)}$ are behaving periodically in time, but at each oscillation
the role of most synchronized population switches from one to the other.

In a limited region above the diagonal and for $g_c > 0.055$ the collective behaviour is still 
symmetric but irregular ({\it Collective Chaos}),
as revelead by the two macroscopic attractors (see Fig. \ref{fig.2}(b)).
Furthermore, in this case one can observe quite wide oscillations of the order
parameters of the two populations in the range $0.4 \le r^{(0)}, r^{(1)} \le 1$,
as shown in Fig. \ref{fig.2}(c). Whenever one population gets synchronized,
with an order parameter $\simeq 1$, the other partially desynchronizes reaching
values $r \simeq 0.4$. These collective oscillations in the order parameters
occur on quite long timescale with respect to the periods of oscillations of
the two macroscopic field $E^{(0)}$ and $E^{(1)}$ reported in Fig. \ref{fig.2}(h).
However, the oscillations in the level of synchronization induce
modulations with periods of the same order in the field dynamics.
In a previous paper~\cite{olmi2010} we have demonstrated that the finite-amplitude Lyapunov exponent~\cite{fsle}, for this state, coincides with the microscopic maximal Lyapunov exponent, 
thus suggesting that the microscopic chaos is induced by the collective drive and therefore the origin of chaos is indeed collective in this case.

Moreover, in a strip above the chaotic region, one can observe a symmetric collective quasiperiodic motion on a {\it Torus} $T^2$ for both populations
(see Fig. \ref{fig.2} (c)). This means that the quasiperiodic motion of the fields is accompanied by a dynamics of the single neurons along a torus $T^3$. 
An analogous regime has been previously reported in~\cite{kura_coll} 
for a population of coupled Stuart-Landau oscillators. Here, we find it in a model where the single units are described by an single variable.
Furthermore, the motion on the macroscopic $T^2$ attractor reported in Fig. \ref{fig.2} (c) can be characterized by estimating the winding numbers for 
various system sizes, we observe that the winding number is constant, indicating that the torus survives in the thermodynamic limit. In this case, we 
observe quite regular antiphase oscillations in the synchronization order parameters between values $0.4 \le r^{(0)}, r^{(1)} \le 0.8$
occurring on time scales definitely longer than those associated to the oscillations of the 
macroscopic fields (as shown in Fig. \ref{fig.2} (f) and (i).

Finally, for yet larger $g_s$-values both populations converge towards a {\it Splay State}, characterized by constant fields, no collective
motion and periodic microsocpic evolution of the neurons. This is not surprising, as we already know that for the chosen 
$\alpha$- and $a$-values, the splay state is stable in a single population of neurons for $G > G_0 \equiv 0.425$~\cite{vvres,zillmer2}.


\section{Diluted Networks}

In order to observe the influence of dilution on the dynamics,
we considered in absence of dilution a PS-FS state with broken symmetry and a Chaotic state.
In particular, we have analyzed the modifications of the macroscopic attractors, of
the level of synchronization, as well as the microscopic dynamics
induced by cutting randomly links for these 2 states.

\begin{figure}
\includegraphics*[angle=0,width=0.5\textwidth]{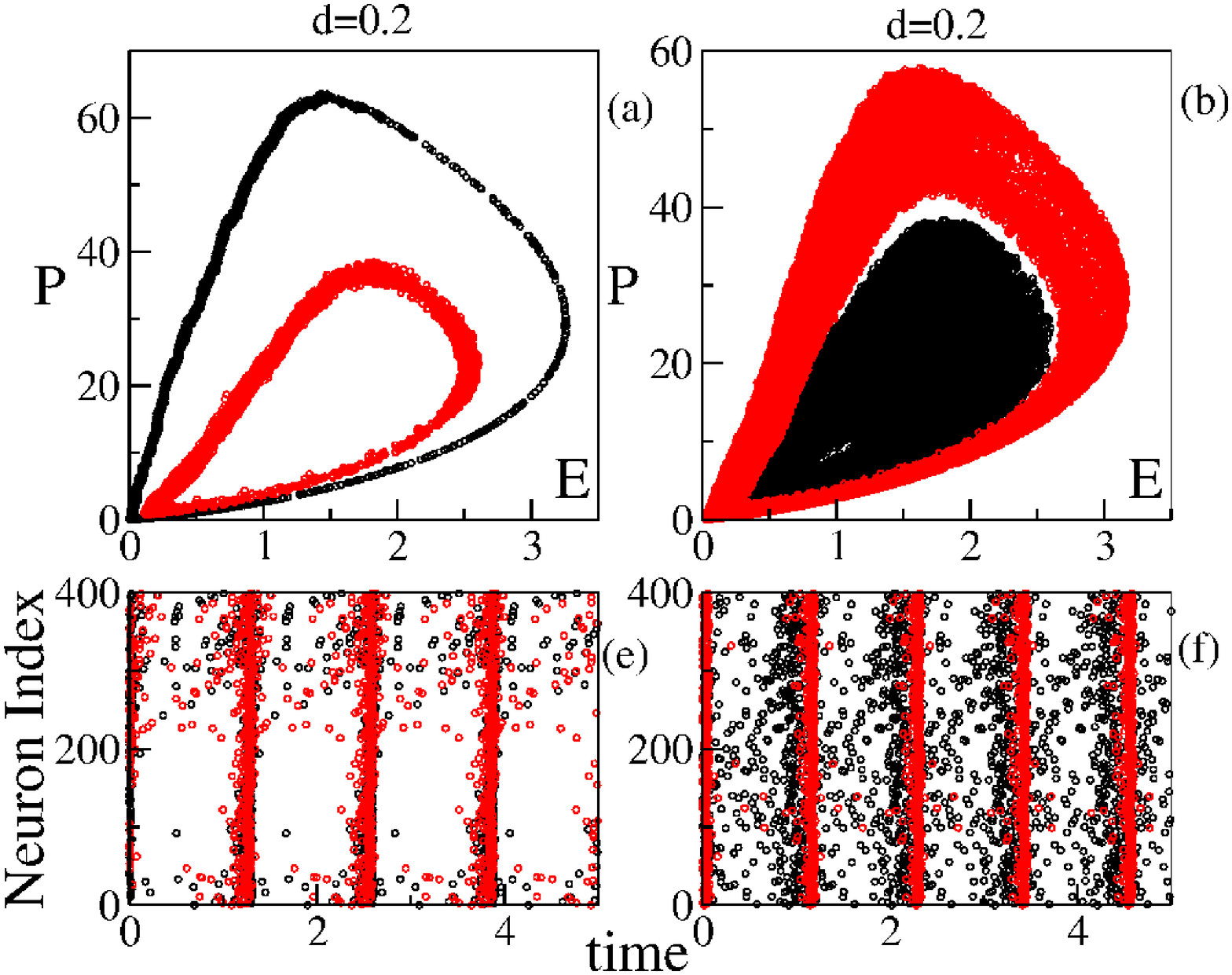}
\includegraphics*[angle=0,width=0.5\textwidth]{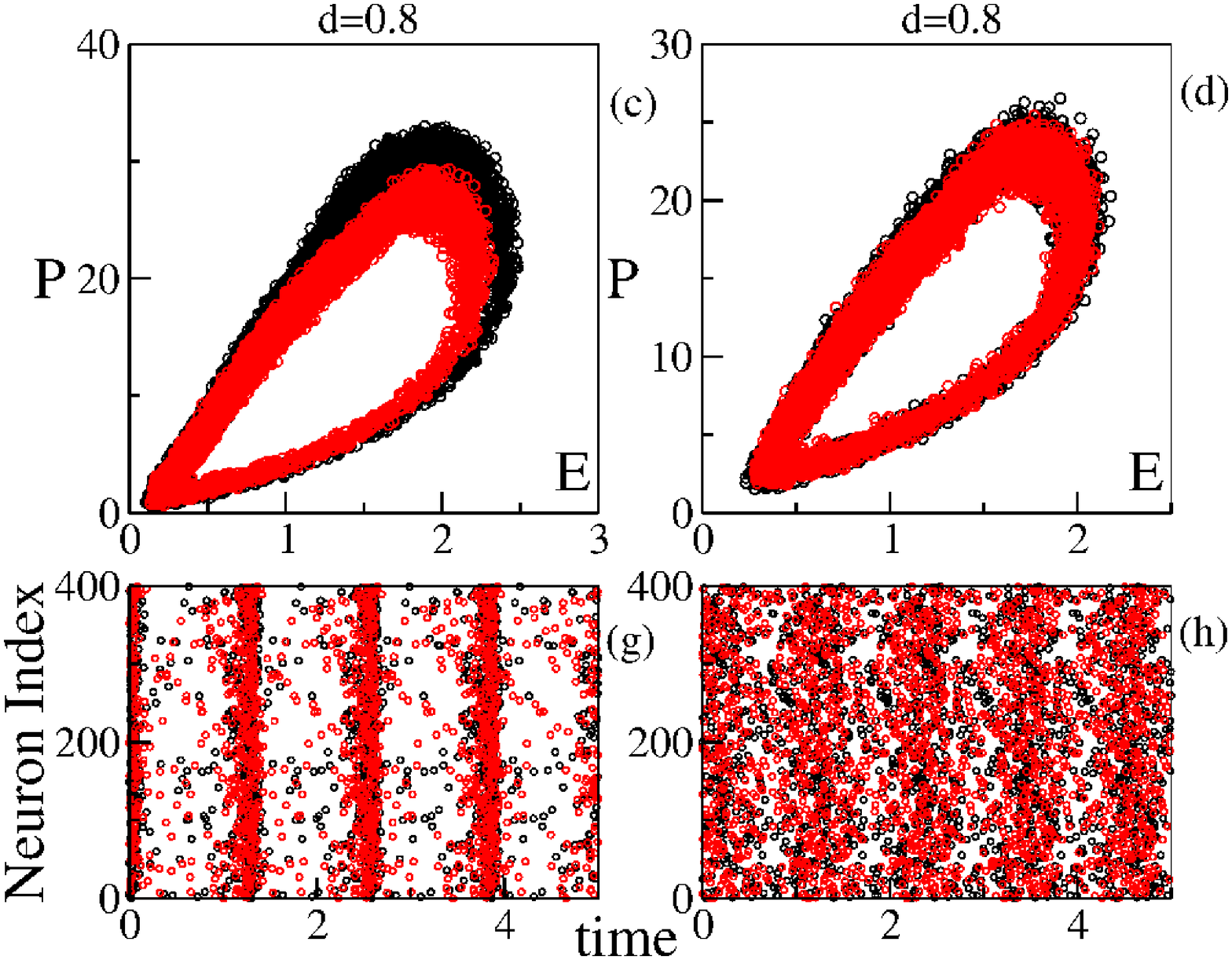}
\includegraphics*[angle=0,width=0.5\textwidth]{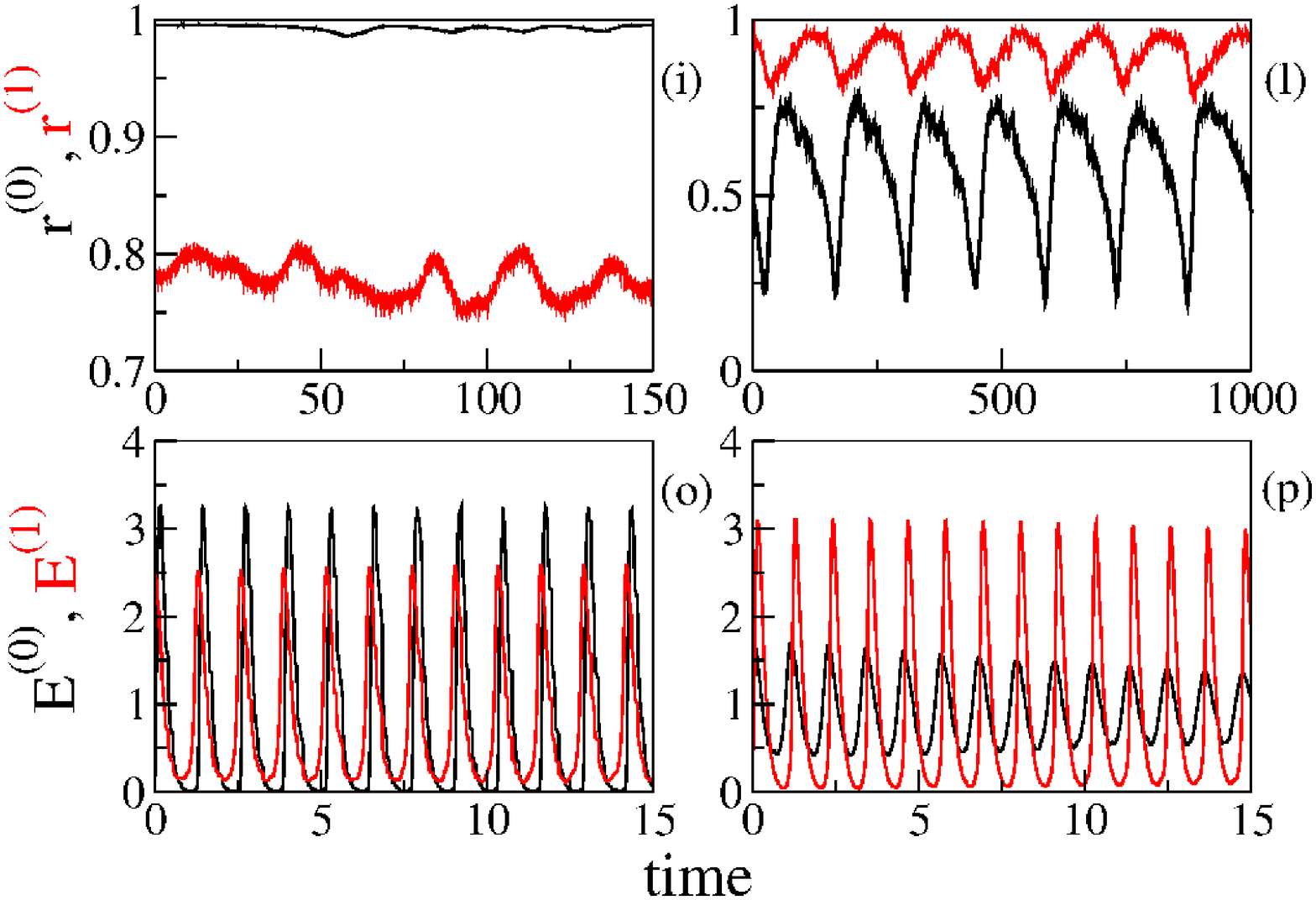}
\includegraphics*[angle=0,width=0.5\textwidth]{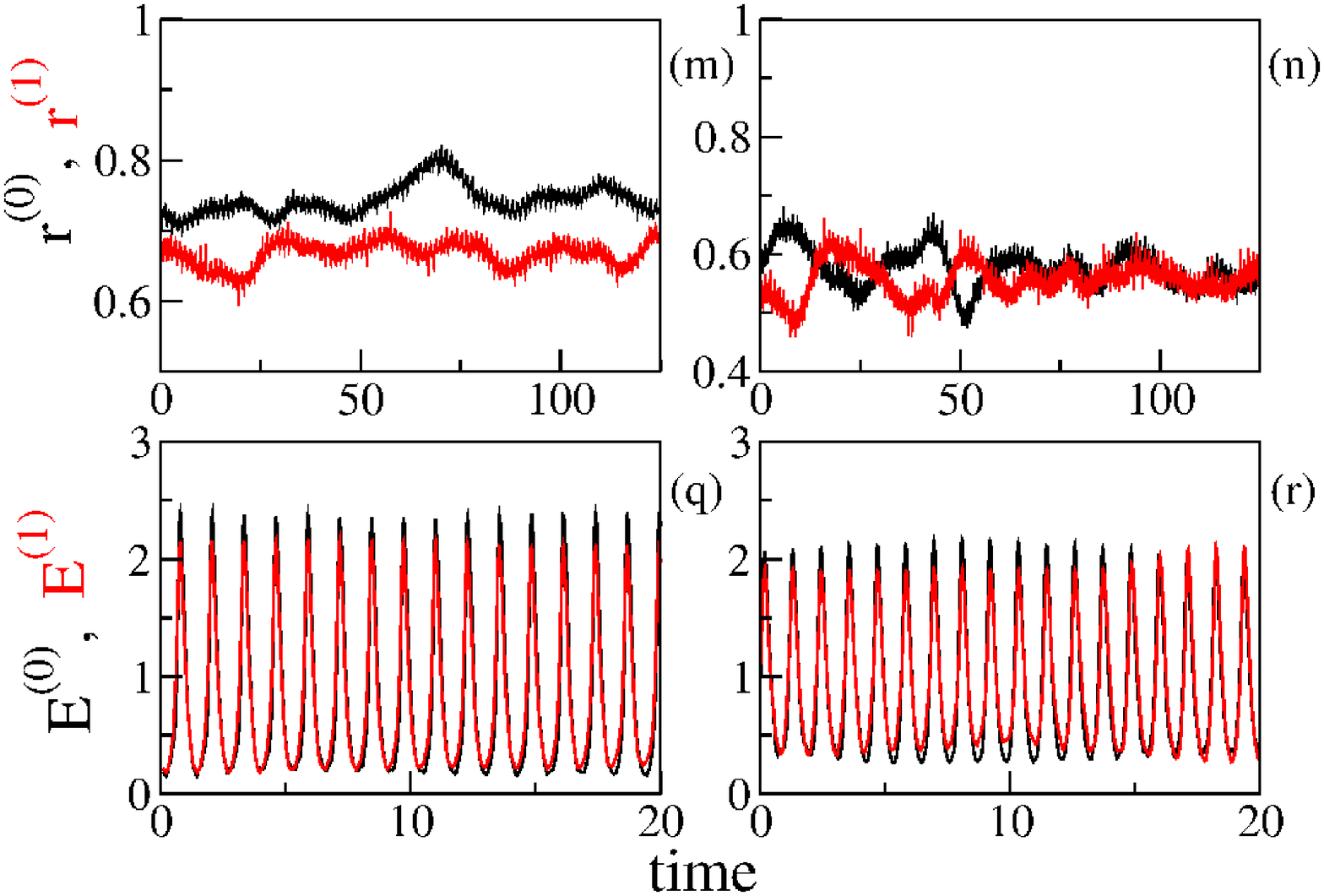}
\caption{(Color Online) {\bf Influence of Dilution.} Macroscopic and
microscopic characterization for two different dilutions (namely$d=0.2$ and 0.8) of two states
that at $d=0$ were FS-PS (first and third columns) and chaotic (second and fourth columns).
In the first row the corresponding macroscopic attractors are reported,
namely, $P \equiv E+\alpha \dot E$ vs $E$ are shown; the raster plots are
shown in the second row; the time evolution of the order parameters $r^{(0)}$ and $r^{(1)}$
is reported in the third row, while that of the macroscopic fields
$E^{(0)}$ and $E^{(1)}$ is shown in the forth row.
The variables corresponding to population 0 (resp. 1) are shown in black (resp. red). 
As regards the parameter values, $(g_c=0.04,g_s=0.1)$ for the first and third columns,
and $(g_c=0.08,g_s=0.16)$ for the second and fourth columns.
The employed values of dilution are reported  over the corresponding columns.
}
\label{fig.3a}
\end{figure}

\begin{figure}
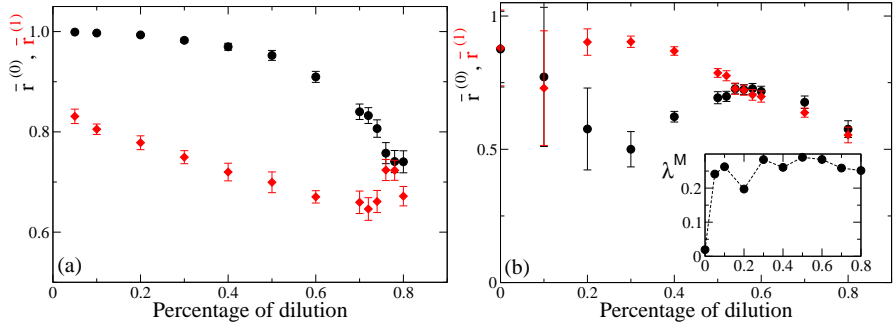

\includegraphics*[angle=0,width=0.5\textwidth]{RvsDilutionGc0.04Gs0.1.eps}
\includegraphics*[angle=0,width=0.5\textwidth]{RvsDilutionGc0.08Gs0.16.eps}
\caption{(Color Online) 
In panels (a), (b) are reported the average values of the order parameters $\bar{r}^{(0)}$ and $\bar{r}^{(1)}$
as a function of the percentage of dilution. In the inset is reported the maximum Lyapunov exponent as a function
of the percentage of dilution. As regards the parameter values, $(g_c=0.04,g_s=0.1)$
in (a) and $(g_c=0.08,g_s=0.16)$ in (b).
}
\label{fig.3b}
\end{figure}

To characterize the two macroscopic states, we have decided to consider the level of 
synchronization in the two populations. In general, we observe that the effect of dilution is, as expected, to reduce the level of synchronization in the system.
In particular, starting from the PS-FS state in the fully coupled case, the average values ${\bar r}^{(0)}$ and ${\bar r}^{(1)}$ 
remain distinct up to some critical dilution $d_c \simeq 0.75$ (as shown in Fig.~\ref{fig.3b} (a)).
For intermediate values of the dilution, in the range $ 0.2 \le d \le d_c$,
a broken symmetry state is still observable, characterized by two periodically oscillating fields with associated different attractors
(see Figs~\ref{fig.3a} (a),(e), and (o)). Therefore, we can safely classify
this as a chimera PS1-PS2 state, despite the dilution induces fluctuations in the 
macroscopic fields~\cite{simo,david1}. For larger dilution, above the critical value, the two 
attractors essentially merge (as shown in Figs~\ref{fig.3a} (i)), but both the macroscopic fields 
are still  presenting clear collective periodic oscillations even at these levels of dilution (see
Figs~\ref{fig.3a} (c)), confirming the robustness of the PS states in this model.

As a general remark when we considered the influence of dilution on a PS1-PS2 state ,
we observed a similar scenario, obviously without an initial window where the FS was
still observable.

The dilution has a quite peculiar effect on the chaotic, symmetric,
state; in fact, up to dilution $d \simeq 0.2$, we did not observe any new effect,
as evidenced by the average value of the synchronization order parameters reported in 
Fig.~\ref{fig.3b} (b). However, already at $d=0.2$ the dilution induce a symmetry break
among the two population dynamics. This is clear in Fig.~\ref{fig.3a} (b), where
one population is still in a collective chaotic state, similar to the one
observed for the globally coupled system, while the other reveals an attractor analogous
to the one seen for the PS state. This is even more evident by considering the time evolution 
of the order parameters, while one population exhibits large oscillations of $r^{(0)}$, similar
to the one observed for the chaotic state, the other reveals more limited oscillations
(see Fig.~\ref{fig.3a} (l)). By further increasing the dilution, the system show a clear 
chimera PS1-PS2 state over a range $ 0.3 \le d \le 0.5$.
For $d > 0.5$ the two attractors merge in a commons PS state, analogously to the previously
considered set of parameters (as shown in Fig.~\ref{fig.3a} (d)(h)(p) and (r) for $d=0.8$).

In this latter case we also measured the maximal Lyapunov exponent $\lambda^M$ and we observed
that it stays positive for all the considered dilution values (see inset Fig. ~\ref{fig.3a}(b)). 
However, while for vanishing dilutions the origin of the chaotic dynamics can be considered
as a collective effect induced by the chaotic motion of the coupled macroscopic fields,
analogous to the chaotic state observed for two fully coupled populations~\cite{olmi2010},
for larger dilution we expect chaotic effects to be present at the level of the single populations,
in the form of (microscopic) {\it weak chaos}. This form of chaos disappears in the thermodynamic
limit, and it is due to stochastic fluctuations of the single macroscopic fields induced 
by finite in-degree effects~\cite{simo,david1}. At intermediate dilution both effects are present
and the level of chaoticity is bigger with respect to the fully coupled case (where, 
$\lambda_M \simeq 0.02$); this is also evident from Fig.~\ref{fig.3a} (b) where one attractor 
appears as being chaotic, while the other is in a PS state plus finite size fluctuations.
To summarize, the system in absence of dilution, thanks to the interaction of the two
populations, exhibits {\it collective chaos}, the dilution induces another form of chaos
termed {\it weak chaos}, because it present only in systems of finite size. However, for
the chosen system size and parameters the level of chaoticity due to finite size fluctuations 
is definitely higher than that due to collective chaos.

\section{Noisy Dynamics}

In this case, we consider as unperturbed state a network with a small level of
dilution $d=0.2$ and we study how the noise modify the original dynamics.
In absence of noise we consider once more a chimera PS-FS state
and a chaotic state. Please, notice that the small dilution modifies the phase diagram 
shown in Fig.\ref{fig.1} for the globally coupled case. In particular in order to observe 
a chaotic symmetric state, we have been forced to employ parameter values slightly 
different from those considered in the previous Section for the same state, namely 
we used $(g_c=0.08,g_s=0.2)$.

\begin{figure}
\includegraphics*[angle=0,width=0.5\textwidth]{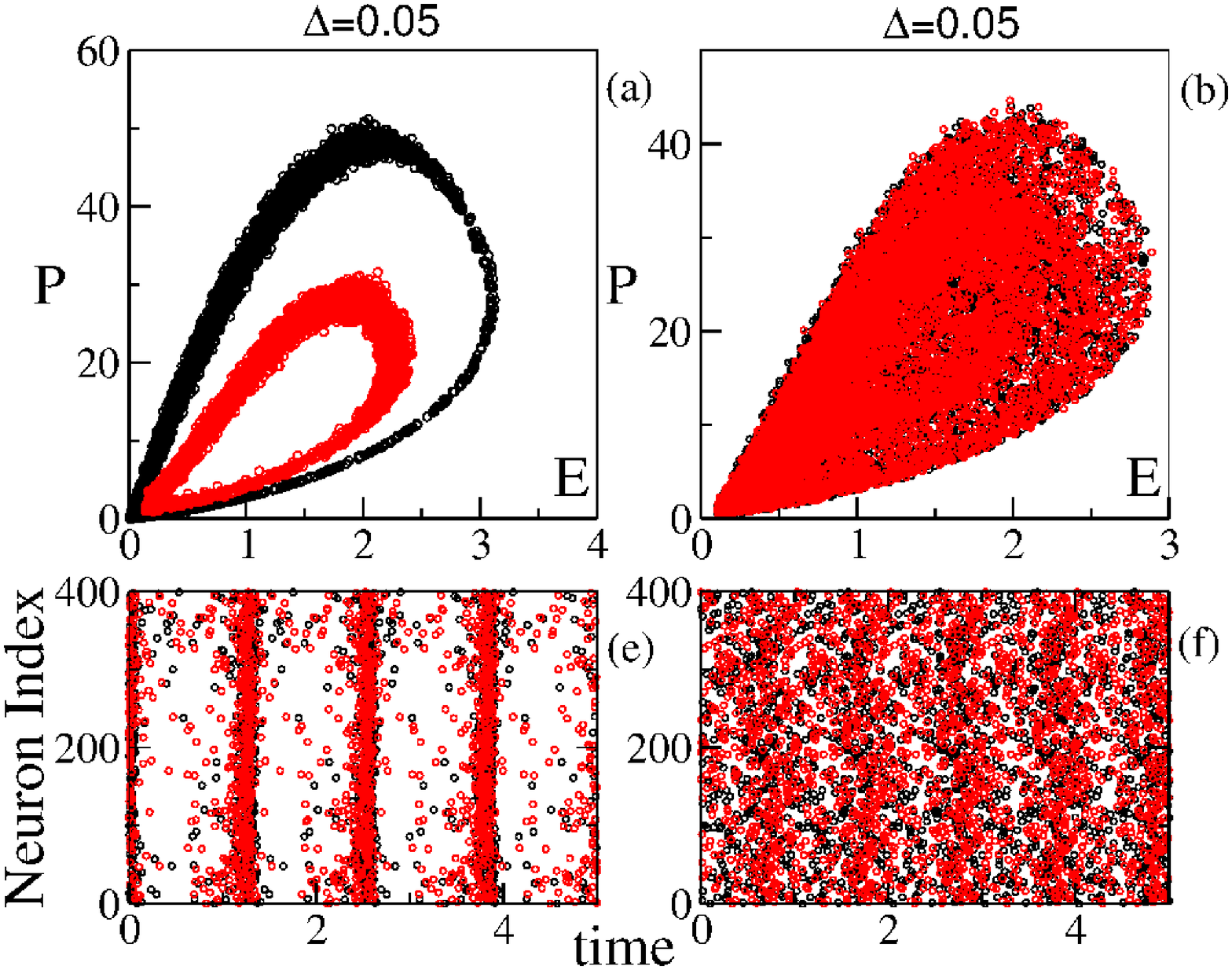}
\includegraphics*[angle=0,width=0.5\textwidth]{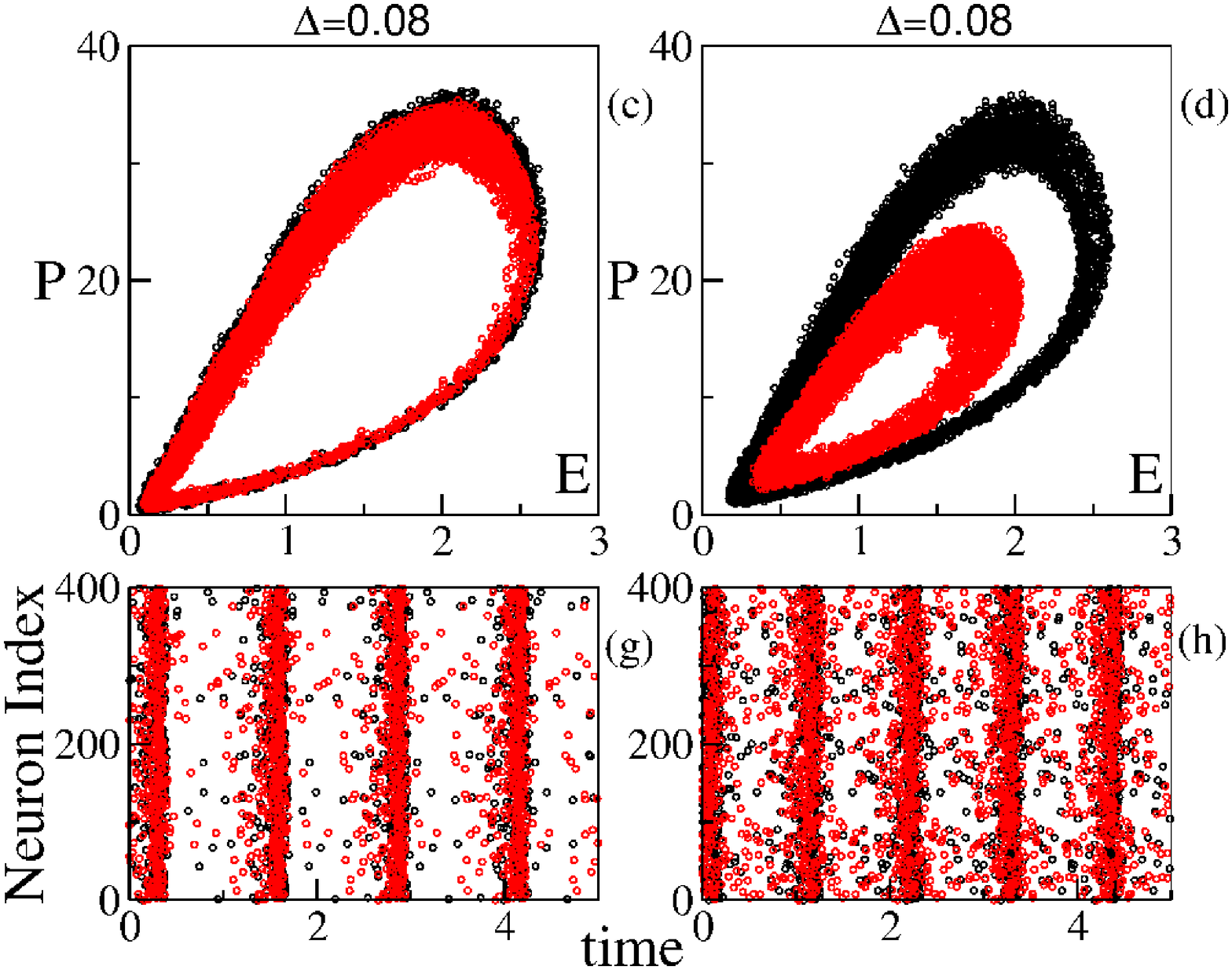}
\includegraphics*[angle=0,width=0.5\textwidth]{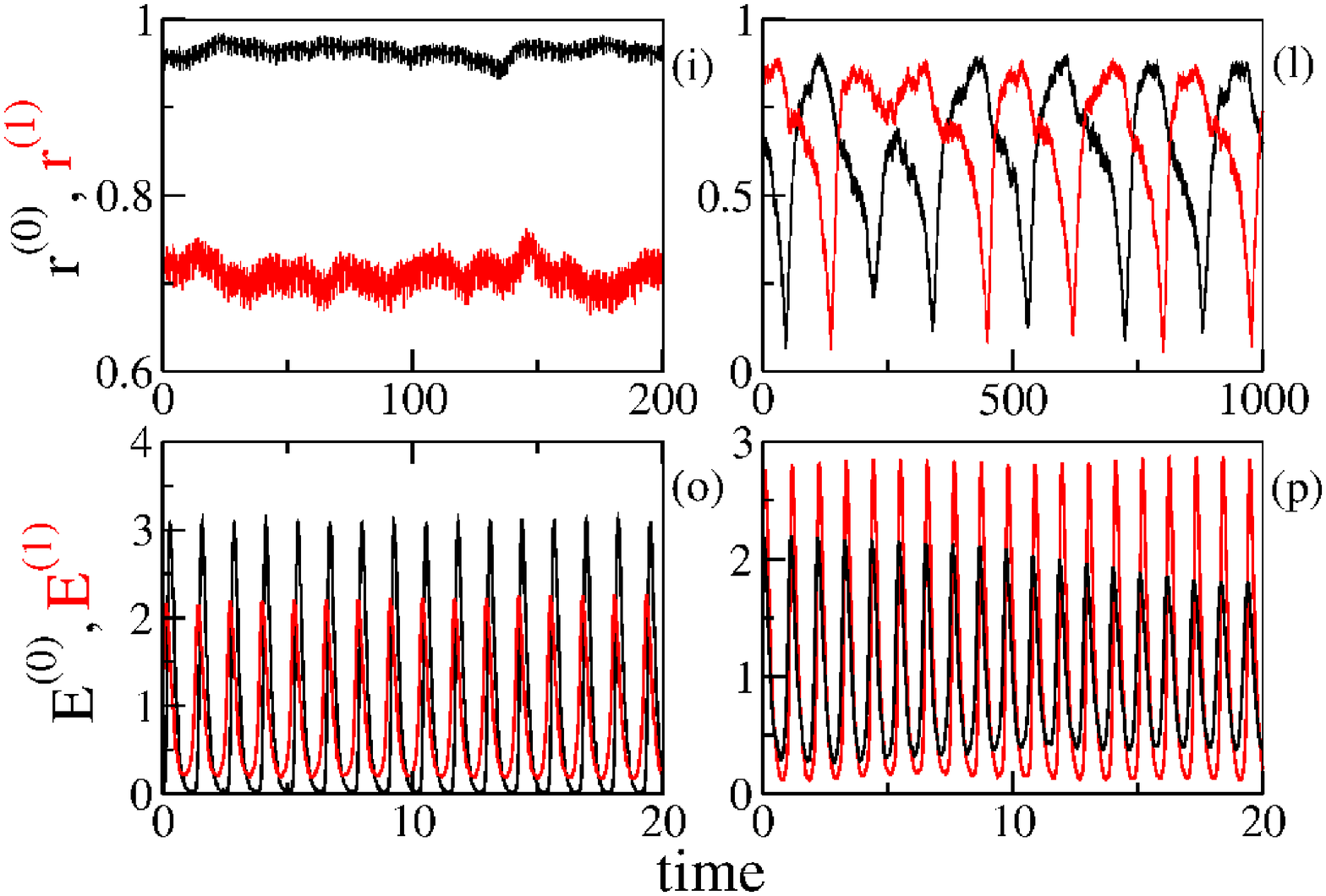}
\includegraphics*[angle=0,width=0.5\textwidth]{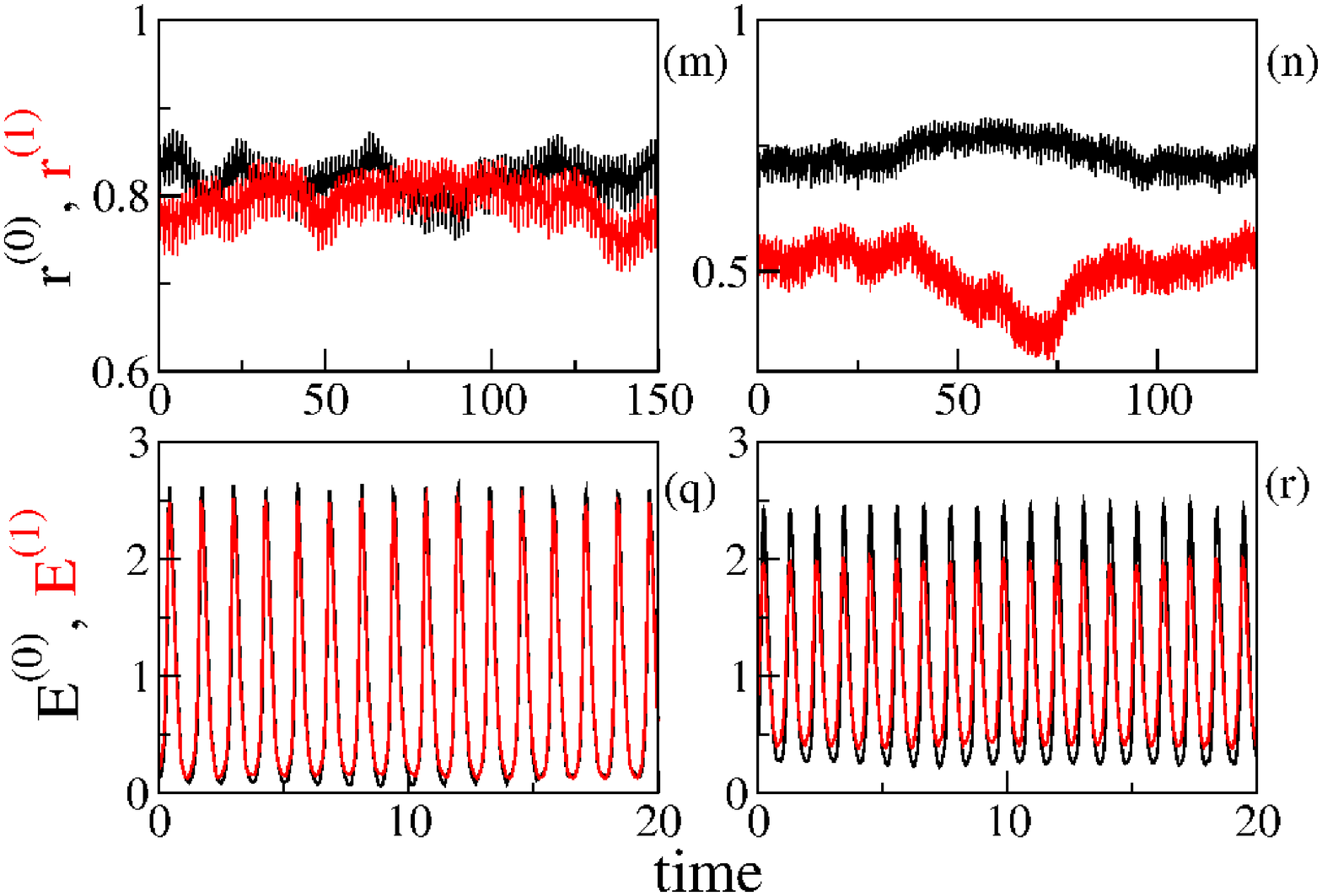}
\caption{(Color Online) {\bf Noise Influence.} 
Macroscopic and microscopic characterization for two different noise amplitudes,
(namely, $\Delta = 0.05$ and $0.08$) of two states
that for $\Delta \to 0$ and $d=0.2$ were FS-PS (first and third columns) and chaotic (second and fourth columns).
In the rows the same variables as in Fig.~\ref{fig.3a} are displayed.
Macroscopic attractors displayed by reporting
The variables corresponding to population 0 (resp. 1) are shown in black (resp. red). As regards the parameter values, $(g_c=0.04,g_s=0.1)$ for first and third columns,
and $(g_c=0.08,g_s=0.2)$ for second and fourth columns.
The employed noise amplitude are reported above the corresponding columns.
In all cases the dilution was fixed to $d=0.2$ and $N=400$.}
\label{fig.4}
\end{figure}

\begin{figure}
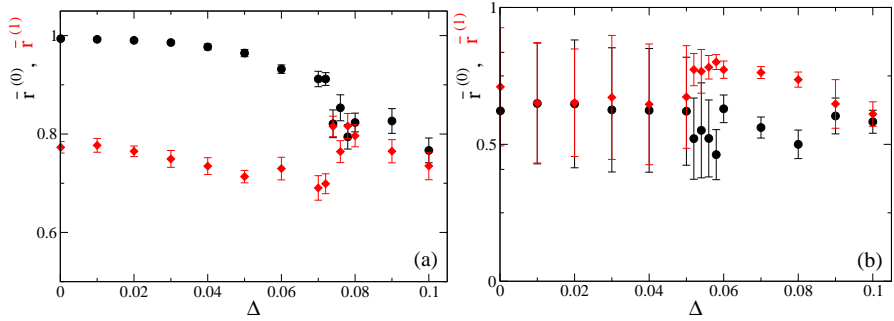

\includegraphics*[angle=0,width=0.5\textwidth]{R_aveVSnoisegc0.04gs0.1.eps}
\includegraphics*[angle=0,width=0.5\textwidth]{R_aveVSnoisegc0.08gs0.2.eps}
\caption{(Color Online) 
In panels (a), (b) are reported the average values of the order parameters $\bar{r}^{(0)}$ and $\bar{r}^{(1)}$
as a function of the noise $\Delta$. As regards the parameter values, $(g_c=0.04,g_s=0.1)$
in (a) and $(g_c=0.08,g_s=0.2)$ in (b). In both cases the dilution is $d=0.2$.
}
\label{fig.5}
\end{figure}

For the chimera FS-PS, it is evident from Fig.~\ref{fig.5} (a) that the 
complete synchronization in one population persist only up to noise of amplitudes
$\Delta \simeq 0.02$, however the two populations behave differently over a quite
wide range of noise amplitudes (namely, $0 \le \Delta \le 0.07$). In all this range
we observe chimera states of the type PS1-PS2, obviously with fluctuations in the
macroscopic variables induced by noise, as it is evident from Fig.~\ref{fig.4} (a),
(e), (i), and (o). By further increasing the noise amplitude a complete
symmetry is recovered but the fields still exhibit periodic collective oscillations
as shown in Fig.~\ref{fig.4} (q). Also all the other indicators suggest that each 
population is still PS, in particular the synchronization degree remains quite
high ${\bar r}^{(0)},{\bar r}^{(1)} \simeq 0.8$ (see also Figs~\ref{fig.4} (c),(g),
(m) and (q)).

For what concerns the chaotic state, this remains symmetric and characterized
by an unique chaotic attractor up to noises of quite large amplitude, namely $\Delta = 0.05$.
As evident, from Figs.~\ref{fig.4} (b),(f),(l) and (p), all the characteristics
of a collectively chaotic state seems present: overlapping chaotic attractors
filling a closed portion of the phase space, anti-phase irregular oscillations 
in the order parameters over long time scales etc. The quite unexpected result
is that by further increasing noise the symmetry of the attractors is broken and
the system evolves towards a chimera PS1-PS2 state, which is observable in the
range $0.06 \le \Delta \le 0.08$ (as shown in Fig.~\ref{fig.5} (b)).
A specific example of this broken symmetry state is reported in
Figs.~\ref{fig.4} (d),(h),(n) and (r) for $\Delta = 0.08$. For
even larger noise amplitudes the two attractors converge towards
a common PS state with level of average synchronization $\bar r \simeq 0.6$.

\section{Discussion}

A first important aspect to notice, is that in the present model the Chimera states FS-PS and PS1-PS2
do not coexist with a stable regime where both populations are FS, as usual
for Chimera states emerging in phase oscillator populations. This implies that in the present case the initial conditions should not be 
prepared in some peculiar way to observe the emergence of broken symmetry states, therefore they are not induced by the choice of the initial 
conditions as in most of the examined models. Spontaneously emerging Chimera, in system where the FS was unstable, have been reported also for chains of Hodgkin-Huxley neurons~\cite{sakaguchi2006} and of Stuart-Landau oscillators~\cite{bord2010}.

As  general results, we observe that dilution or noise have a 
similar influence on the studied macroscopic dynamics, despite
random dilution represents a quenched form of disorder, while
dynamical noise an annealed one. In particular,
starting from a broken symmetry state dilution or noise reduce the level
of synchronization in the two population, leading the dynamics of the two
networks to be more and more similar for increasing dilution/noise. 
On the other hand, starting from a symmetric state, namely a chaotic
one, the role of disorder is to break (at some intermediate dilution 
or noise amplitude) the symmetry among the dynamics of the two populations. 
Thus in this case, the disorder can promote the emergence of a chimera-like
state (a PS1-PS2) in a range of parameters where the dynamics was
fully symmetric in the globally coupled deterministic set-up.
For large dilution/noise the system always ends up in a partially 
synchronized regime. This can be explained by the fact that
the stable state, for the chosen parameter and for identical coupling among neurons 
of both populations (namely, $g_s = g_c$), is the regime PS.
Indeed, for large dilution or noise the heterogeneity in the synaptic coupling among 
neurons lying in one population or in another become less pronounced and the PS emerge.
Theferore, disorder has at some intermediate level a constructive effect inducing the birth 
of a more complex (broken symmetry) state from a fully symmetric one,
similarly to what reported in~\cite{semenova}, where {\it coherence-resonance chimeras}
have been observed.

Another interesting aspect, is that the chimera-like states PS1-PS2 are
quite robust to dilution, they can be observed up to $80 \%$ of randomly
broken links within each population, while previous results on 
phase oscillators pointed out that chimeras are observable up to $8 \%$
of dilution~\cite{laing2012}. The origin of this noticeably difference
is probably due to the fact that in this model PS states can be observed even in sparse networks
with an extremely small in-degree ( $K \simeq 10$ ) as shown in~\cite{luccioli2012}.
Furthermore, another stabilizing factor is the choice of the cross synaptic current,
in our model the effect of one population on the neurons of the other population
is mimicked via a macroscopic mean field, representing the average synaptic
current.

The reported results represent only a first step in the study of the emergence of chimera states
in neural systems characterized by a sparse topology and by the presence of noise.
Further analysis will be required to investigate more realistic models and
to understand if chimera states can have a role in the encoding of information 
at a population level in brain circuits.

\begin{acknowledgement}
We thank E.A. Martens, A. Politi and S. Gupta for extremely useful interactions.
This article is part of the research activity of the Advanced Study Group 2016
{\it From Microscopic to Collective Dynamics in Neural Circuits} performed
at Max Planck Institute for the Physics of Complex Systems in
Dresden (Germany).
 
\end{acknowledgement}
\section*{Appendix: Accurate event driven map for the two fully coupled populations}
\addcontentsline{toc}{section}{Appendix}
%

In the two symmetrically fully coupled populations setup here discussed
we find various kind of symmetric and symmetry broken states; in particular we find synchronized
states. The integration of such states can be become a difficult issue, due to numerical
round-off it can become extremely difficult to determine the next firing neuron, thus
to increase the numerical accuracy and to avoid spurious clustering due to numerical round-off, 
we implemented the following integration scheme.

In particular, instead of integrating the membrane potentials, we performed the integration of
the logarithm of the difference of two successive neurons. This transformation is uniquely defined
in globally coupled systems, since the order of the neurons passing threshold is 
preserved in time. Therefore, it is possible to define an ordered list of the potentials and, on this basis,
to define uniquely the ``neighbours'' of a neuron. Given a set of $N$ membrane potentials 
$\left\lbrace x_j^{(k)} \right\rbrace_{j=1,\ldots,N}$, with $k=0,1$ depending on the considered
family, we introduce at a generic time $t$ the following $N+1$ auxiliary variables:
\begin{eqnarray} \nonumber
 \omega_1^{(k)}(t)&=&\ln\left[1- x_1^{(k)}(t)\right] \\ \nonumber
 \omega_j^{(k)}(t)&=& \ln\left[x_{j-1}^{(k)}(t)- x_j^{(k)}(t)\right] 
 \qquad j=2,\ldots,N
 \\ \nonumber
 \omega_{N+1}^{(k)}(t)&=&\ln\left[x_N^{(k)}(t)\right]
\end{eqnarray}
where the threshold (resp. reset) value is $x_{th}=1$ (resp. $x_R=0$) and $x_1^{(k)}$ is the next to
threshold neuron. 

Since we would like to define an event driven map for the two coupled families, it is necessary to
find which neuron is going to fire next and then evolve the membrane potentials of the two populations
untile the successive spike emission.
The evolution of the two populations is different and it depends on the fact that the firing neuron
belongs to the considered family or not. Let us schematize the algorithm in three steps:
\begin{enumerate}
 \item As a first step we compare $x_1^{(0)}$ with $x_1^{(1)}$ to identify to which family
the firing neuron belongs.
\item As a second step we check if the firing neuron belongs to a family which has already
fired at the previous event or not. Depending on this, we have two possible alternatives:
if the next and previous firing neurons belong to the same family we iterate the network 
as in point (a) below, otherwise as in point (b).
\begin{enumerate}
 \item Let us suppose that the firing population is the family $(0)$.  
 We evolve all the $N+1$ variables $\left\lbrace \omega_j^{(0)} \right\rbrace_{j=1,\ldots,N+1}$  
 for the firing family, while, for the other family, it is sufficient to evolve just the variables
 $\omega_1^{(1)}$ and $\omega_{N+1}^{(1)}$.
 All the above mentioned variables are evolved for a lapse of time corresponding to the interval
 elapsed from the last firing event of the family $(0)$. 
 \item If the firing family is $(1)$ and previously fired a neuron of family $(0)$, the
 evolution is more complicated. The variables $\omega_1^{(k)}$
 and  $\omega_{N+1}^{(k)}$ of both families are integrated for 
 the time interval elapsed from the last firing time of family $(0)$.
 The $N-1$ variables $\left\lbrace \omega_j^{(1)} \right\rbrace_{j=2,\ldots,N}$ should
 be instead evolved for a longer time corresponding to the last interspike interval 
 associated to family $(1)$, because these variables have not been updated since 
 the last firing of family $(1)$.      
\end{enumerate}
 \item The firing family is iterated in the comoving frame: this amounts to update the membrane 
 potentials and to shift the index of all neurons by one unit. 
 The membrane potentials of the other family are updated in the fixed reference frame.
 \item The simulation is iterated by repeating the above three steps.
\end{enumerate}

In order to evolve the a linearized system  the previous algorithm is no more effective
since now it is necessary to evolve all the variables at each time step in order to calculate
the linearized equations in the tangent space. In this case we use directly the 
difference of the membrane potentials of two successive neurons instead of the logarithm.
We still search for the first to fire neuron between the two populations and we treat 
differently the variables of the two populations at each time step depending on which
neuron has emitted a spike previously. As in the previous case we employ different reference
frames for the firing or not firing family (see point 3 above).

%
%
%

\biblstarthook{References may be \textit{cited} in the text either by number (preferred) or by author/year.
\footnote{Make sure that all references from the list are cited in the text. Those not cited should be moved to a separate \textit{Further Reading} section or chapter.} 
The reference list should ideally be \textit{sorted} in alphabetical order -- even if reference numbers are used for the their citation in the text. 
If there are several works by the same author, the following order should be used: 
\begin{enumerate}
\item all works by the author alone, ordered chronologically by year of publication
\item all works by the author with a coauthor, ordered alphabetically by coauthor
\item all works by the author with several coauthors, ordered chronologically by year of publication.
\end{enumerate}
The \textit{styling} of references\footnote{Always use the standard abbreviation of a journal's name according to the ISSN \textit{List of Title Word Abbreviations}, 
see \url{http://www.issn.org/en/node/344}} depends on the subject of your book:
\begin{itemize}
\item The \textit{two} recommended styles for references in books on \textit{mathematical, physical, statistical and computer sciences} 
are depicted in ~\cite{science-contrib, science-online, science-mono, science-journal, science-DOI} and ~\cite{phys-online, phys-mono, phys-journal, phys-DOI, phys-contrib}.
\item Examples of the most commonly used reference style in books on \textit{Psychology, Social Sciences} are~\cite{psysoc-mono, psysoc-online,psysoc-journal, psysoc-contrib, psysoc-DOI}.
\item Examples for references in books on \textit{Humanities, Linguistics, Philosophy} are~\cite{humlinphil-journal, humlinphil-contrib, humlinphil-mono, humlinphil-online, humlinphil-DOI}.
\item Examples of the basic Springer style used in publications on a wide range of subjects such as \textit{Computer Science, Economics, Engineering, Geosciences, Life Sciences, Medicine, Biomedicine} are ~\cite{basic-contrib, basic-online, basic-journal, basic-DOI, basic-mono}. 
\end{itemize}
}

%
%
%
%
%
%
%

\end{document}